# Meso-scale modeling: beyond local equilibrium assumption for multiphase flow

Wei Wang*, Yanpei Chen
State Key Laboratory of Multiphase Complex Systems, Institute of Process Engineering, Chinese Academy of Sciences, Beijing 100190, China
*Email: wangwei@ipe.ac.cn

**Abstract**

This is a summary of the article with the same title, accepted for publication in Advances in Chemical Engineering, 47: 193-277 (2015). Gas–solid fluidization is a typical nonlinear nonequilibrium system with multiscale structure. In particular, the mesoscale structure in terms of bubbles or clusters, which can be characterized by nonequilibrium features in terms of bimodal velocity distribution, energy nonequipartition, and correlated density fluctuations, is the critical factor. Traditional two-fluid model (TFM) and relevant closures depend on local equilibrium and homogeneous distribution assumptions, and fail to predict the dynamic, nonequilibrium phenomena in circulating fluidized beds even with fine-grid resolution. In contrast, the mesoscale modeling, as exemplified by the energy-minimization multiscale (EMMS) model, is consistent with the nonequilibrium features in multiphase flows. Thus, the structure-dependent multifluid model conservation equations with the EMMS-based mesoscale modeling greatly improve the prediction accuracy in terms of flow, mass transfer, and reactions as well as the understanding of flow regime transitions. Such discrepancies raise the question of the applicability of the local equilibrium assumption underlying the TFM and further shed light to the necessity of mesoscale modeling.

**Introduction**

With increase of gas agitation, a fluidized bed experiences flow states from homogeneous expansion with minimum fluidization, to heterogeneous bubbling fluidization, slugging fluidization, turbulent fluidization, circulating fluidization (provided that a separator is equipped to circulate particles entrained from the top outlet) and pneumatic conveying (Kunii and Levenspiel, 1991; Grace et al., 2006). A fluidized bed is a typical nonequilibrium system because it is time-dependent and not ideally homogeneous. The meso-scale structures in terms of bubbles or clusters are determined by both collective motion of large quantity of particles and interstitial gas eddies of large quantity of molecules.

In contrast to molecular gas flow, granular materials and rapid granular flows (or, granular gas) in fluidization are characterized by the dissipative nature, which leads to lack of both spatial and temporal scale separation (Campbell, 1990). Specifically, at least three typical features are related with nonequilibrium granular flow systems: i) energy unequipartition or nonexistence of one granular temperature; ii) non-Gaussian, bimodal distribution of velocity and iii) strong correlated density fluctuations with bimodal distribution of solids holdup, with respect to voids and clusters. If we incorporate the effects of structures in a fluidized bed with a dilute-dense, structure-based meso-scale modeling, we may update the currently popular two-fluid model (TFM), which is based on the local equilibrium assumption, to structure-dependent multifluid modeling (SFM) by satisfying nonequipartition of energy (two granular temperatures), bimodal velocity distribution (non-Gaussian) and inherent clustering. That approach seems to be a reasonable simplification to the complex multiphase flow in gas-fluidized beds.

**Mesoscale Modeling**

Bearing in mind the nonequilibrium features of granular flows with bimodal velocity and density distributions, we proposed a set of structure-dependent multifluid modeling (SFM) methods (Hong et al, 2012, 2013; Liu et al, 2015; Song et al, 2014). The bimodal velocity distribution was incorporated into the Boltzmann equation, thus introducing the mesoscale structure into the conservation equations for the mass, momentum, energy, and species. Based on these conservations, we deduced and analyzed the EMMS model (Li and Kwauk, 1994) for the steady state of fluidized beds and its stability condition. A generalized framework of the multiscale CFD can thus be established (Wang et al., 2010) with stability constrained characterization of mesoscale structure, as shown in Fig. 1.

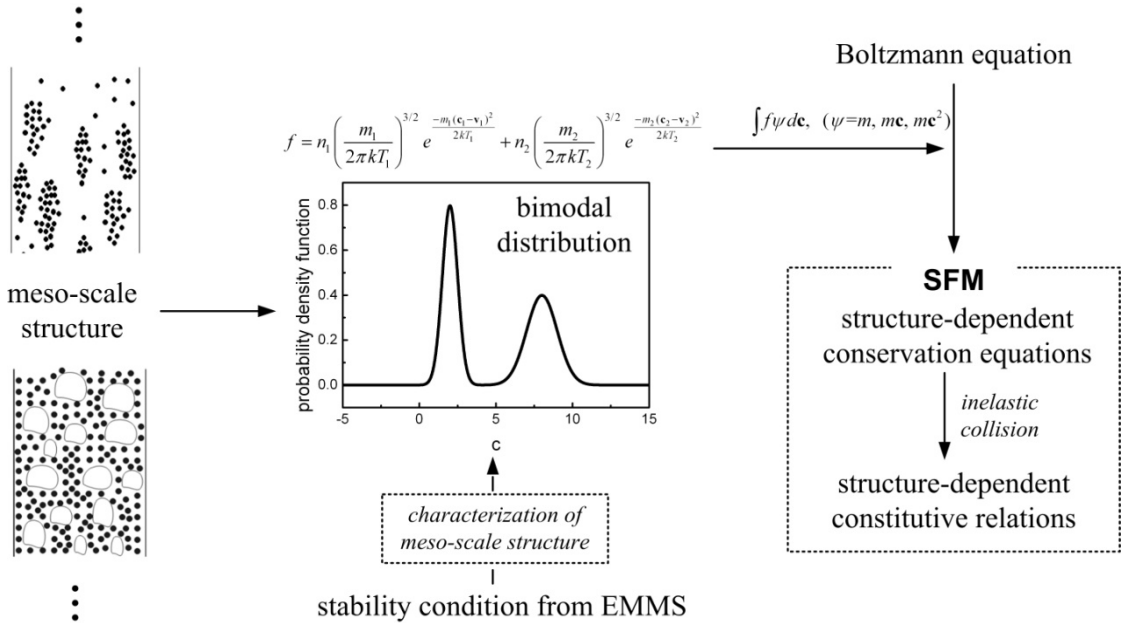

Fig. 1 Schematic of the meso-scale modeling for multiphase flow by introducing bimodal velocity distribution which is beyond local equilibrium

Different from the two-fluid or multi-fluid model where both the gas and solids are treated as fully interpenetrating continua, the four continua in the SFM are not fully interpenetrating. Thus for certain pair of continua, there may be no interphase forces. For example, there is no drag between dilute-phase particles and dense-phase gas, but there are drags between gas and particles in both the dilute and dense phases (for the sake of simplicity, only drag force is discussed in this work, as the other interphase forces, e.g., virtual mass force and lift force, are neglected). The conservation equations for four structure-based continua can be derived following the Eulerian spatial averaging method. Accordingly, it is the closure of the meso-scale drag force that may be used to distinguish different SFMs. Compared to TFM, the SFM differs in its formulation of stress, drag and diffusion stress by including the effect of the dilute-dense structure. SFM may revert to the cluster-based EMMS or bubble-based EMMS hydrodynamic models under different assumptions of structure parameters. Thus SFM unifies these models (Song et al., 2014; Hong et al., 2013).

The quantification of the meso-scale structure in these equations can be empirical, using experimental correlations for the cluster/bubble diameter, or, more fundamentally, from theoretical analysis. According to the principle of compromise in competition (Li and Kwauk, 1994; Li and Huang, 2014), this dependency is determined by the minimization of energy consumption rate for suspending and transporting particles in a fluidized bed. That is, the energy consumption for suspending and transporting particles per unit mass of particles has a tendency to reach its minimum value, Nst→min, through a compromise between the minimum tendencies of the voidage and the energy consumption for suspending and transporting particles per unit volume.

**Disputes in drag closure**

Integration of the normal and shear stresses over the surface of particles gives various interphase forces, in which the drag force represents the time-independent component of the longitudinal force (along direction of relative velocity), besides the time-dependent longitudinal forces (e.g., added mass force, Basset force), transversal forces (or lift forces perpendicular to relative velocity), and the buoyancy irrelevant to slip (Crowe and Michaelides, 2006). Disputes never cease as to the closure of the key component, the effective drag force, when many interacting particles exist in dense multiphase flows. Here, the effective drag force is termed because the symmetrical, steady-state presumptions in defining the drag force for single particles are hard to be satisfied for many particle systems. And the effective drag force for many particles is actually an averaged interphase force defined on the macroscopically steady state, including probably contributions of all the above components of forces owing to tortuous, asymmetrical flow field between particles and interstitial fluid.

There are several possible approaches to close the effective drag force for multiphase flow systems: (1) Assuming drag force equal to effective gravity (say, gravity of particles minus buoyancy): this force balance relation is normally assumed at the reactor scale under steady state, but it is not valid transiently due to local acceleration of particles; (2) Taking the drag coefficient obtained under the conditions of static array or homogeneous suspension of particles: Ergun (1952) and Richardson and Zaki (1954) drag coefficients from experiments and Koch and Sangani (1999) drag coefficient from simulation are widely cited examples, which actually assumes unchanged structure irrespective of the variation of relative velocity; (3) Deriving the drag coefficient by performing a set of fine-grid simulation over a periodic domain and then from the force balance between gravity and drag force: the filtered drag coefficients of Agrawal et al (2001), Igci et al (2008), and Igci and Sundaresan (2011) are typical examples, which take into account the effects of dynamic structure on the drag coefficient but with constraint of local force balance (also local equilibrium); and (4) Deriving the drag coefficient with the local resolution of the energy-minimization multiscale (EMMS) model: the EMMS/matrix drag (Hong et al, 2012; Lu et al, 2009; Wang and Li, 2007) is a typical example, which considers the effects of dynamic structure with nonequilibrium feature as mentioned above.

Of all these approaches, the first one is seldom applied because it contradicts the obvious fact of local acceleration of particles. The second one is widely used in CFD simulation, especially for bubbling fluidized beds, which is expected to be homogeneous in local space and generally satisfy the local equilibrium. Both the third one and the last one have considered the effects of dynamic structure. However, as the third one is obtained under the constraint of force balance over periodic domain, the voidage is the only independent parameter to determine the flow structure properties and thus the structure-dependent drag correction is a function only of voidage. Whereas the EMMS/matrix drag is obtained without such constraint, the structure-dependent drag correction is hence a function of both voidage and slip velocity, which better represents the nonequilibrium features of multiphase flow. More details on the discussion of the functional dependence of drag correction are referred in Wang et al (2010).

**Comparison between methods with/without meso-scale modeling**

To show the advantage of meso-scale modeling over the approaches based on local equilibrium and homogeneous closures, in particular, the TFM, we performed comparison starting from a simple one-dimensional force balance analysis, aiming to shed light on which kind of drag can predict the bi-stable flow state (Ullah et al., 2013a). Then, CFD simulation with and without meso-scale modeling of drag are compared, in particular for the fluidization of fine particles classified as Geldart A (Hong et al., 2016). The effects of meso-scale modeling on both Eulerian-Eulerian (TFM vs EFM) and Eulerian- Lagrangian (MP-PIC) approaches and reactive simulation are discussed. Finally the flow regime map with meso-scale modeling is highlighted to help understand the choking and flooding phenomena (Wang and Chen, 2015).

As to the bistable state analysis on 1D force balance, if we incorporate a homogeneous drag into the force balance equations, only one intersection can be found on the two curves of the drift flux diagram for concurrent-up riser flow, no matter how the operating conditions are varied. If we use the EMMS drag instead, we can find two intersections under certain set of operating conditions, which corresponds to the dilute-dense coexisting flow state, or the choking transition (Ullah et al, 2013a). Further we test the performance of meso-scale modeling under fine-grid resolution. For a bubbling fluidized bed, or a circulating fluidized bed with coarse particles (say, Geldart B, D particles), which seems to be close to the requirement of local equilibriums states, the fine-grid TFM simulation with homogeneous drag is a plausible approach. However, it is not sufficient to resolve all the meso-scale structures in high-velocity fluidized beds and hence the solids flux predicted is not reliable. In contrast, the EMMS-based modelling depends on bi-modal distribution, which is closer to reality and far from local equilibrium, and hence enables reasonable prediction of both axial profiles and solids flux (Hong et al., 2016). Further, as to the reactive simulation of ozone decomposition in a circulating fluidized bed with mesoscale modeling, we found that the results agree with experimental data. If we use the TFM with homogeneous drag and mass transfer coefficient, as indicated in our previous work and review (Dong et al., 2008a; 2008b; Wang et al., 2011), the reaction rate will be greatly overpredicted, resulting in quick, nearly complete conversion of ozone near the distributor. Finally we also provide some hints on how the meso-scale modeling modify or improve the widely applied flow regime maps for fluidization, for which the details are referred to (Ullah et al., 2013b).

## SUMMARY AND PROSPECTS

Mesoscale structure is the core of research for multiphase flows and, more generally, multiphase reaction engineering. Without mesoscale structure, there will be no essential difference between flow/reactions in a lab-scale flask and in a homogeneously distributed, industrial-scale multiphase flow reactor. In other words, there will be no scale-up effects for such multiphase systems. Recognizing the importance of mesoscale structure, chemical engineers have been focusing on it ever since the emergence of modern chemical engineering (fluidization research development is a good example for it), though at that time, there is no such terminology of "mesoscale." Until recently after several decades of efforts, chemical engineering researchers revealed that the same mechanism termed the principle of "compromise in competition" may be obeyed by various kinds of mesoscale structures (Li and Huang, 2014; Li et al, 2013). Though all these progress have been made, we are still far from quantitative understanding of the mesoscale structure due to its nonlinear nonequilibrium nature.

It has been said that "scientists tackle those problems which can be solved; engineers are faced with problems which must be solved" (Sherwood et al, 1975). Such a statement well describes the difference between chemical engineering and mechanics communities when encountering nonequilibrium multiphase flows. The past efforts on fluid mechanics modeling of fluidization largely stays on the hydrodynamic description of fluid-like granular flows, which is based on the local equilibrium assumption, seeming analogy between granular gas and molecular gas and successful application of Navier–Stokes equations for single-phase flows. Though the mesoscale structure (or termed as microstructure) has been recognized to greatly affect the flow state of rapid granular flow or granular gas (Campbell, 1990; Goldhirsch, 2003), the clear statement of breakdown of hydrodynamics for inelastic particles was made until Du et al (1995). Even now, the TFM still prevails in the simulation of fluidization.

Recent years have witnessed a clear tendency of fusion between the communities of granular matter physics, fluid mechanics, and fluidization engineering, due to more and more understanding of the nonequilibrium features inherent in granular flows and fluidization, e.g., non-Gaussian velocity distribution, energy nonequipartition, and correlated density fluctuations, and hence a blossom of mesoscale modeling research from various aspects of the problems and angles of view. In these efforts, the EMMS-based, SFM has found successful applications by greatly improving the prediction accuracy in terms of flow, mass transfer, and reactions as well as the understanding of flow regime transitions. In contrast, traditional TFM, which depends on local equilibrium and homogeneous distribution assumptions, fails to predict the dynamic, nonequilibrium phenomena in CFBs even with fine-grid resolution.

We can expect the disputes over the applicability of the local equilibrium assumption that underlies the TFM will remain for a longer time, as the secrets of scale-dependent, nonequilibrium multiphase flows are not fully understood. More in-depth investigation should be conducted with close coupling among experimental measurement, theoretical derivation,

and numerical simulations:

- In experiments, microscopic, noninvasive observation of the velocity distribution, correlated density fluctuation, or clustering will be helpful to quantify the collective behavior of granular flows under different heating or energy driving conditions.
- With experimental quantification of mesoscale structures, statistical analysis based on nonequilibrium distribution may require novel mathematical skills to unravel the complex dependence of stress–strain relation on the mass exchange between phases. The structure-dependent energy analysis may help elucidate the dependence between energy dissipation and structural parameters, and how to relate such structural-dependent analysis and the nonlinear nonequilibrium thermodynamics remains a challenge, especially for the scale-dependent granular flow or fluidization systems.
- Direct numerical simulation (DNS) is a powerful tool to investigate the velocity distribution and density fluctuation of multiphase flow systems, thus facilitates revealing the mechanisms of nonequilibrium behavior. Due to its high demand in computing resources, current DNS is largely limited to simulations over static arrays of particles or small-sized, periodic flow domains, which are expected to be close to local equilibrium states. Thus, the nonequilibrium characteristics of multiphase flow are hard to be fully revealed. Recent release of hybrid computing hardware boosts the rapid development of DNS with respect to the scales of time and space that allows us a more realistic DNS of a fluidized bed and more in-depth analysis of nonequilibrium characteristics. Such a big jump of capability may radically modify our research mode, bring us to the new paradigm of virtual process engineering, and help explore the mesoscience on the horizon (Li, 2015).

**References**

Agrawal K, Loezos PN, Syamlal M, Sundaresan S: The role of meso-scale structures in rapid gas-solid flows, J Fluid Mech 445:151–185, 2001.

Campbell, C.S., 1990. Rapid granular flows. Annual Review of Fluid Mechanics 22, 57-92.

Crowe CT, Michaelides EE: Basic concepts and definitions. In Crowe CT, editor: Multiphase flow handbook, Boca Raton, 2006, CRC Press.

Dong, W., Wang, W., Li, J., 2008a. A multiscale mass transfer model for gas-solid riser flows: Part 1 - Sub-grid model and simple tests. Chemical Engineering Science 63, 2798-2810.

Dong, W., Wang, W., Li, J., 2008b. A multiscale mass transfer model for gas-solid riser flows: Part II - Sub-grid simulation of ozone decomposition. Chemical Engineering Science 63, 2811-2823.

Du, Y.S., Li, H., Kadanoff, L.P., 1995. Breakdown of hydrodynamics in a one-dimensional system of inelastic particles. Physical Review Letters 74, 1268-1271.

Ergun, S., 1952. Fluid flow through packed columns. Chemical Engineering Progress 48, 89-94.

Ge, W., Li, J., 2002. Physical mapping of fluidization regimes - the EMMS approach. Chemical Engineering Science 57, 3993-4004.

Goldhirsch, I., 2003. Rapid granular flows. Annual Review of Fluid Mechanics 35, 267-293.

Grace, J. R., Leckner, B., Zhu, J., Cheng, Y., 2006. Fluidized beds. In: (Crowe, C.T. Ed.) Multiphase Flow Handbook. CRC Press, Boca Raton.

Hong K, Chen S, Wang W, Li J: Fine-grid two-fluid modeling of fluidization of Geldart A particles, Powder Technol , 2016.

Hong, K., Shi, Z., Wang, W., Li, J., 2013. A structure-dependent multi-fluid model (SFM) for heterogeneous gas-solid flow. Chemical Engineering Science 99, 191-202.

Hong, K., Wang, W., Zhou, Q., Wang, J., Li, J., 2012. An EMMS-based multi-fluid model (EFM) for heterogeneous gas-solid riser flows: Part I. Formulation of structure-dependent conservation equatio Hong, K., Chen, S., Wang, W., Li, J., 2015. Fine-grid two-fluid modeling of fluidization of Geldart A particles. Powder Technology (in press)ns. Chemical Engineering Science 75, 376-389.

Igci Y, Andrews AT, Sundaresan S, Pannala S, O'Brien T: Filtered two-fluid models for fluidized gas-particle suspensions, AIChE J 54:1431–1448, 2008.

Igci, Y., Sundaresan, S., 2011. Constitutive models for filtered two-fluid models of fluidized gas-particle flows. Industrial & Engineering Chemistry Research 50, 13190-13201.

Kunii, D., Levenspiel, O., 1991. Fluidization Engineering. Butterworth-Heinemann, Boston

Koch DL, Sangani AS: Particle pressure and marginal stability limits for a homogeneous monodisperse gas fluidized bed: kinetic theory and numerical calculations, J Fluid Mech 400:229–263, 1999. Fluid Mech 400:229–263, 1999.

Li, J., Huang, W., 2014. Towards Mesoscience: The Principle of Compromise in Competition. Springer.

Li, J., Kwauk, M., 1994. Particle-Fluid Two-Phase Flow: Energy-Minimization Multi-Scale Method. Metallurgy Industry Press, Beijing.

Li, J., 2015. Approaching virtual process engineering with exploring mesoscience. Chemical Engineering Journal 278, 541-555.

Li J, Ge W, Wang W, et al: From multiscale modeling to meso-science, Berlin Heidelberg, 2013,
Springer.

Liu, C., Wang, W., Zhang, N., Li, J., 2015. Structure-dependent multi-fluid model for mass transfer and reactions in gas-solid fluidized beds. Chemical Engineering Science 122, 114-129.

Lu B, Wang W, Li J: Searching for a mesh-independent sub-grid model for CFD simulation of gas-solid riser flows, Chem Eng Sci 64:3437–3447, 2009.

Richardson J, Zaki W: Fluidization and sedimentation—part I, Trans Inst Chem Eng 32:38–58, 1954.

Sherwood, T.K., Pigford, R.L., Wilke, C.R., 1975. Mass Transfer. McGraw-Hill, New York.

Song, F., Wang, W., Hong, K., Li, J., 2014. Unification of EMMS and TFM: structure-dependent analysis of mass, momentum and energy conservation. Chemical Engineering Science 120, 112-116.

Ullah A, Wang W, Li J: Evaluation of drag models for cocurrent and countercurrent gas-solid flows, Chem Eng Sci 92:89–104, 2013a.

Ullah, A., Wang, W., Li, J., 2013b. "Generalized fluidization" revisited. Industrial & Engineering Chemistry Research 52, 11319-11332.

Wang, W., Li, J., 2007. Simulation of gas-solid two-phase flow by a multi-scale CFD approach - Extension of the EMMS model to the sub-grid level. Chemical Engineering Science 62, 208-231.

Wang, W., Lu, B., Zhang, N., Shi, Z., Li, J., 2010. A review of multiscale CFD for gas-solid CFB modeling. Int. J. Multiphase Flow 36, 109–118.

Wang, W., Ge, W., Yang, N., Li, J., 2011. Meso-scale modeling—The key to multi-scale CFD simulation. Advances in Chemical Engineering 40, 1-58.

Wang, W., Chen, Y., 2015. Mesoscale modeling: beyond local equilibrium assumption for multiphase flow. Advances in Chemical Engineering 47, 193-277.